# Dynamic Control of Nonlinear Emission by Exciton-Photon Coupling in WS$_2$ Metasurfaces

Mudassar Nauman,[1,2] Domenico de Ceglia,[3] Jingshi Yan,[2] Lujun Huang,[4] Mohsen Rahmani,[5] Costantino De Angelis,[3] Andrey E. Miroshnichenko,[4] Yuerui Lu,[1*] and Dragomir Neshev[2*]

[1]School of Engineering, Australian National University, Canberra, ACT, 2601, Australia

[2]ARC Centre of Excellence for Transformative Meta-Optical Systems, Department of Electronic Materials Engineering, Research School of Physics, The Australian National University, Canberra, ACT, 2601, Australia

[3]University of Brescia, Department of Information Engineering, Via Branze 38, 25123, Brescia, Italy

[4]School of Engineering and Information Technology, University of New South Wales, Canberra, ACT, 2600, Australia

[5]Advanced Optics and Photonics Laboratory, Department of Engineering, School of Science and Technology, Nottingham Trent University, Nottingham, NG11 8NS, UK

*Corresponding author. Email: dragomir.neshev@anu.edu.au (D. N.), yuerui.lu@anu.edu.au (Y. L.)

**Abstract**

Transition metal dichalcogenides (TMDCs) have demonstrated significant potential as versatile quantum materials for light absorption and emission. Their unique properties are primarily governed by exciton-photon interactions, which can be substantially enhanced through coupling with resonant photonic structures. For example, nonlinear light emission, such as second harmonic generation (SHG) is doubly enhanced when the incident wave is resonant simultaneously with the excitonic and photonic resonance. However, the excitonic absorption of incident waves can significantly dump the SHG emission. Here, we propose and demonstrate a tunable enhancement of SHG by leveraging virtual coupling effects between quasi-bound states in the continuum (qBIC) optical resonances and tunable excitons in arrays of high-index WS$_2$ crescent metaatoms. These crescent metaatoms excites a pure magnetic type qBIC resonance, enabling dynamic control and enhancement of nonlinear optical processes in visible spectrum. Our findings demonstrate that an array of WS$_2$ crescent metaatoms, exhibiting qBIC resonance at half the exciton energy, enhances SHG efficiency by more than

98-fold compared to monolayer WS$_2$ (1L-WS$_2$) and four orders of magnitude relative to unpatterned WS$_2$ film. This substantial SHG enhancement is tunable as a function of temperature and polarization angle of incident light, allowing us to obtain control of the virtual coupling and SHG efficiency in the visible spectrum (600-650 nm). Our work opens new avenues toward next-generation reconfigurable meta-optics devices.

1. **Introduction**

Strong nonlinearities at the nanoscale, such as in subwavelength structures and metasurfaces (1-4, are of paramount importance for applications in nonlinear and quantum optics (5-8). Low-loss, high-index dielectric metasurfaces made of Si (9), Ge (10) and III-V semiconductors (11, 12) are often the materials of choice for enhancing the nonlinear response at the nanoscale. However, most of the demonstrations reported to date have been focused on passive devices (7-16). Following the current need for industrial applications such as dynamic nonlinear holography, imaging (13), bioelectronics (17) and sensing (18), the new paradigm in the field of dielectric metasurfaces is to switch to active/tunable metasurfaces (19-21) and trigger dynamic control of the nonlinear emission.

Transition metal dichalcogenides (TMDCs) offer an important alternative to conventional semiconductors due to their unique properties, such as high refractive index, novel weak van der Waals (vdWs) stiction forces, and atomically confined excitons even in bulk films (22-24). These properties offer new opportunities for the realization of tunable dielectric metasurfaces. For example, firstly, the high refractive index and weak vdWs stiction forces can be exploited to transform a flat bulk TMDC film into a deep subwavelength array of meta-atoms on any desired transparent substrate. Second, via external stimuli such as environmental index (25), strain (26), electrical gating (27), electrical and temperature treatment (28, 29), the atomically confined excitons can be spectrally tuned over large spectral widths. In TMDCs, excitons exhibit exceptionally large binding energies, enabling their stability at room temperature and even higher, in stark contrast to III-V semiconductors where excitons are only stable at cryogenic temperatures. The interaction of these excitons with light can be further enhanced by the presence of photonic resonances, such as in optical metasurfaces. The higher the quality factor (Q-factor) of the resonances, the stronger the exciton-photon coupling is. As such, metasurfaces exhibiting qBIC resonances have attracted strong recent interest (5,13,30), including for enhancing nonlinear emission (5). To date, however, the tunable nature of excitons has been exploited mainly in atomically thin TMDC films to realize tunable linear

(26,27) and nonlinear (28,31,32) photonic devices. The few works exploiting bulk excitons have focused on the strong coupling regime, where the incident is fully cycled between the excitons and the cavity, resulting in mode splitting (33-35). However, the absorption of the incident light is detrimental to nonlinear processes, such as the upconversion in the SHG process. To avoid the single-photon absorption of the excitation, the excitonic resonance should not be at the same energy as the qBIC resonance. To date, this regime has never been exploited in TMDC metasurfaces for the tunability and enhancement of nonlinear emissions. This research gap and the above-mentioned findings raise the intriguing question of whether excitonic resonances in bulk TMDCs can be harnessed to underpin next-generation reconfigurable flat photonic devices.

In this work, we propose and demonstrate a tunable single-crystalline homogeneous subwavelength $WS_2$ metasurface that supports high Q resonance qBIC at a fundamental wavelength of 1220 nm. As the bulk $WS_2$ has excitonic resonance in the visible spectrum around 610 nm. Hence, a metasurface with qBIC at 1220 nm gives us an exciting opportunity to engineer the novel physics (at the quantum level) between exciton (610 nm) and second harmonic photons of qBIC at room temperature. Such strong excitonic-photonic quantum interaction in the $WS_2$ metasurface enormously enhances the SHG by more than 98-folds compared to 1L-$WS_2$ and 4 orders of magnitude compared to unpatterned $WS_2$ film. Intriguingly, the SHG enhancement can be tuned through external stimuli, such as temperature (which regulates A-type exciton ($E_o^A$)), and pump polarisation angle ($\varphi$), which controls the qBIC excitation. Moreover, experimental studies and theoretical modelling have been done, and remarkable SHG enhancement is observed in the visible spectrum (600-650 nm), where $WS_2$ is opaque. Our design opens entirely new strategies for highly tunable single crystalline nonlinear photonic devices.

## 2. Results

### 2.1. Design of qBIC WS₂ metasurface for tunable SHG

In TMDCs, owing to their weak vdWs stiction forces between layers, many salient features of monolayer and bilayers continue to exist even in the bulk form (*36*). These include spin layer locking (37) and intralayer A-type excitons ($E_o^A$). The SHG in TMDCs is intrinsically enhanced when the incident or generated photons are in resonance with the exciton (28,31,32,38). However, it is beneficial for the exciton to be in resonance with the two-photon energy of the incident light, to reduce the excitonic absorption of the strong pump. As the SHG process is quadratic to the incident power, strong photonic resonances can be further employed to boost the SHG efficiency. In this way, the excitonic $E_o^A$ and photonic qBIC resonances can interfere through a virtual energy level, resulting in the multiplication of the total SHG enhancement. We note that because of the inversion symmetry in the bulk TMDC, the SHG is emitted only from the top and bottom layers (39), which can interfere in the far-field. Intriguingly, likewise, in monolayer form, the energy level of $E_o^A$ in bulk can be tuned as a function of a gated electric field (28) or temperature (27,32), thereby opening a unique avenue to engineer the position and the strength of the SHG generation in TMDCs. The concept of this tunable spectral and energy enhancement is schematically depicted in the three-level energy diagram in Fig. 1A.

In our work, we chose WS$_2$ as a bulk TMDC metasurface due to the convenience of spectrally matching the WS$_2$ excitonic resonance with the available femtosecond lasers. However, the concept is applicable to other TMDC materials. In the proposed scheme, at room temperature (RT), the qBIC and $E_o^A$ of bulk WS$_2$ metasurface are in resonance with the fundamental and SH waves simultaneously to enormously boost the SHG from the WS$_2$ metasurface. Whilst change in temperature (from $-100$ to $100°C$) spectrally tune the $E_o^A$ (from 600 to 625 nm corresponding to energy levels of 2.06 to 1.99 eV, respectively). Such spectral shift of $E_o^A$ (as a function of temperature) breaks the resonance condition, thereby significantly reducing the SHG. Because of the low thermo-optic coefficient of WS$_2$, the change in temperature (from $-100$ to $100°C$) doesn't alter the spectral position of qBIC. This property gives us a degree of freedom to manipulate the SHG by spectrally tuning the $E_o^A$ only excitonic energy near the SH wavelength as a function of temperature.

Along with temperature tunability, we consider pump polarisation and wavelength-dependent SHG from WS$_2$ qBIC metasurface, as shown in the schematic Fig. 1B. The proposed metasurface has a thickness, $t = 220$ nm, periodicity, $P = 600$ nm, bottom, and top radius $r = 220$ nm and 180 nm, respectively. The parameters $\delta$ and $\delta''$ shown in Supplementary Materials Fig. S4 represents the radius and axial displacement of the cylindrical cavity carved into one side of each metaatom. This modification breaks the metasurface symmetry and controls the coupling of the qBIC mode to free space radiation as a function of pump polarisation angle ($\varphi_{pump}$) at normal incidence (40). By tuning the pump wavelength, the SHG is enhanced proportionally to the square of the quality factor (Q-factor) of the qBIC resonance. However, a giant enhancement together with unidirectional emission of the SHG occurs when

$\lambda_{qBIC} = 2 \times \lambda_{exciton}$ and $\varphi_{pump}$ is along the one of the major axes ($\varphi_{pump} = 90°$) of metaatoms, as shown in Fig. 1B.

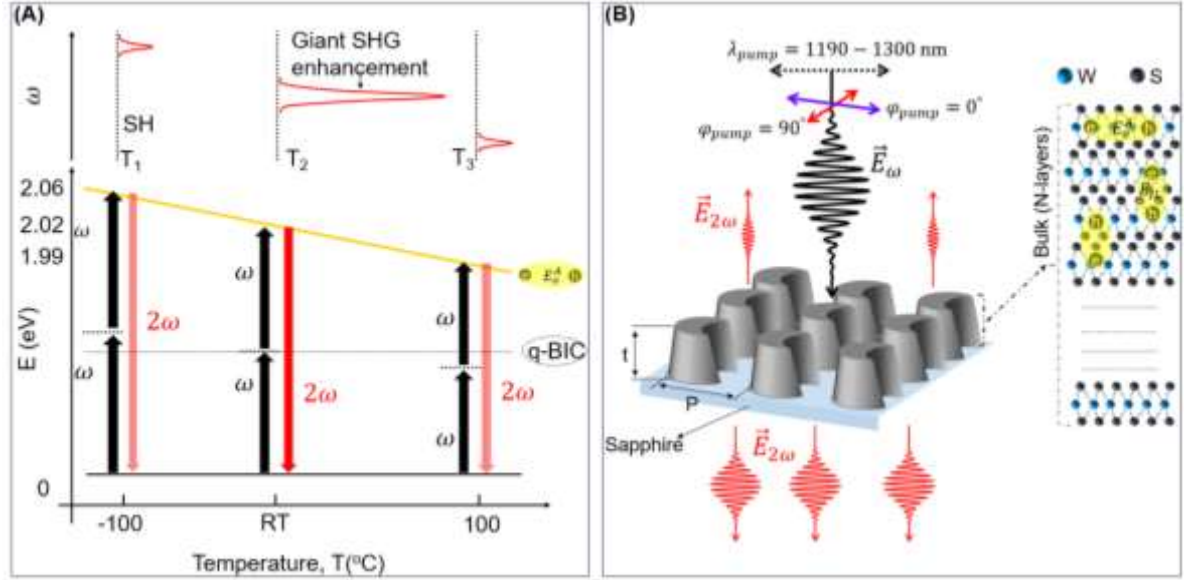

**Fig. 1. Principle of tunable giant enhancement in SHG.** (A) Three-level model showing the thermally tunable energy level of exciton and its overlapping with the SH of qBIC. (B) Schematic illustration of nonlinear $WS_2$ qBIC metasurface having qBIC at 1220 nm (half the energy level of A-type exciton $E_o^A$), the parameters P (periodicity), and t (thickness). Bulk $WS_2$ is a layered material, and each layer has confined $E_o^A$, as shown in the inset. $\vec{E}_\omega$ and $\vec{E}_{2\omega}$ represent the electric field vectors for fundamental and SH frequency, respectively. The SHG enhancement tunable via pump polarisation ($\varphi_{pump}$), which controls the qBIC excitation.

### 2.2. Fabrication and linear response

To maintain the pristine quality, a large area bulk $WS_2$ film of thickness 220 nm has been mechanically exfoliated and dry transferred onto a sapphire substrate (~500 μm thick). Then , a flat bulk film is carved into subwavelength arrays of crescent metaatoms using a standard electron beam lithography (EBL) method and dry etching. The detail of the fabrication method can be found in the Supplementary Materials Fig. S1. We patterned three metasurfaces, A, B, and C, on the same flake, as illsutrated in Supplementary Materials Fig. S2. The three metasurfaces under study share identical geometric dimensions, except for variations in the central position of a cylindrical feature carved into one side of each metaatom. This modification transforms the cone-shaped metaatoms into crescent-shaped metaatoms. In metasurface A, a cylindrical cavity with a radius of 150 nm and axial displacement of 200 nm along radial axis is introduced on one side of the cone-shaped metaatoms, creating an asymmetric crescent shape, as depicted in Fig. 1B and Supplementary Materials Fig. S4. This asymmetry induces a pure magnetic type qBIC resonance, whose spectral position can be finely tuned by adjusting either the radius of the cylindrical cavity or its axial displacement. The three metasurfaces exhibit strong qBICs at 1220 nm, 1270 nm, and 1305 nm, respectively. As illustrated in Fig. 2A, a pure magnetic type qBIC can be induced in the individual meta-atoms by suppressing the electric dipole through the excitation of an anapole mode at the same spectral position as the magnetic dipole (41).

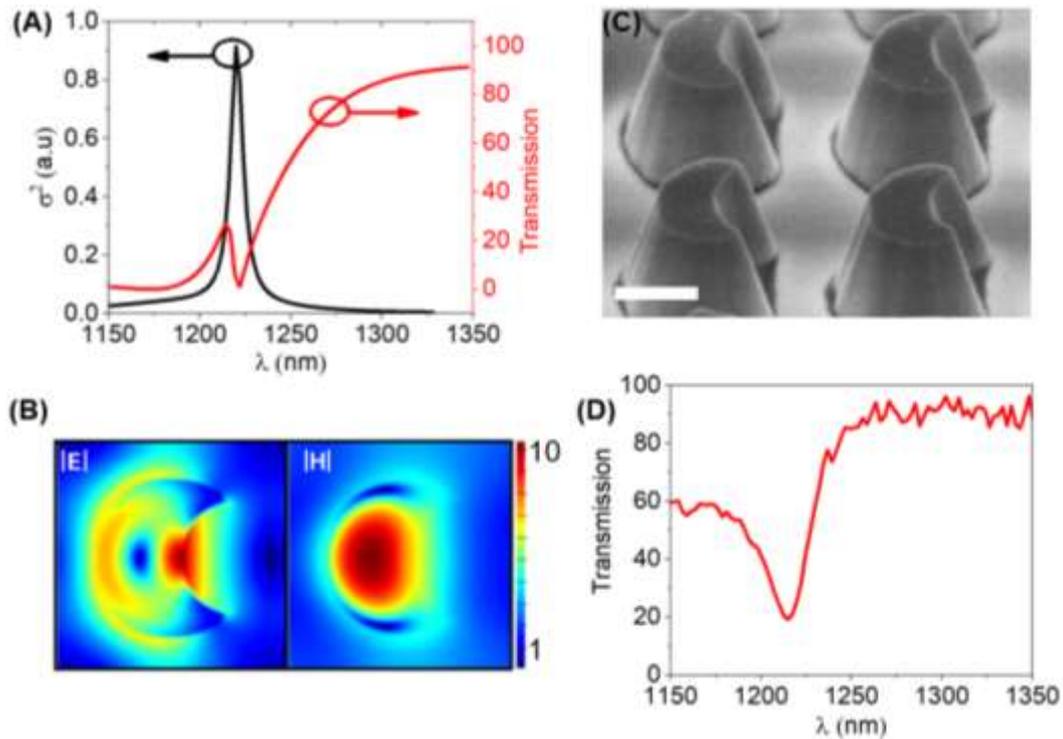

**Fig. 2. Experimental and numerical analysis of the WS$_2$ metasurface.** (A) Multipole expansion of metasurface A, where $\sigma$ represents the effective cross-section and simulated linear spectrum, showing qBIC resonance at 1220 nm. (B) Calculated electric and magnetic field distributions at the resonant $\lambda$ of 1220 nm. (C) SEM image of fabricated WS$_2$ metasurface A on a sapphire substrate. This is our targeted metasurface because it has qBIC at 1220 nm (double the $\lambda$ of $E_o^A$). The scale bar is 500 nm. (D) Measured linear spectrum of metasurface A.

The simulated linear transmission (red curve) and effective scattering cross-section, $\sigma$ (black curve) of metasurface A (targetted metasurface), for which qBIC is at 1220 nm, are shown in Fig. 2A. The electric and magnetic field distributions are shown in Fig. 2B, demonstrating electric field circulation and magnetic field enhancement, as expected for a vertical magnetic dipole mode. The radiation of this mode into the far field is controlled by the asymmetry parameter ($\delta''$). Inriguingly, this asymmetry introduces polarisation sensitivity (asymmetric coupling of MD qBIC resonance with incident light polarisation angle) and enables tunable resonance properties for advanced optical applications, as discussed in detail in Supplementary Materials section 2.3.

As WS$_2$ metaatom is a non-magnetic material, therefore in order to identify the nature of induced qBIC, we performed multipole decomposition, as shown in Fig. 2A. The spherical electrical dipole (ED) has zero response right at the spectral position where magnetic dipole (MD) has the strongest response, as depicted in (Supplementary Materials Fig. S4). The zero response of spherical ED is a clear indication of the crossing point of Cartesian ED and toroidal dipole (TD), resulting in the excitation of anapole mode at the MD resonance, 'the pure MD scattering.' To enhance the second-order nonlinear response, this exotic ideal MD type qBIC has not been exploited/excited so far in other recently reported nonmagnetic III-V semiconductor metasurfaces (42,43). The unfolding nature of the strongly enhanced magnetic field inside the resonator, as illustrated in Fig. 2B, further predicts the nature of qBIC (magnetic dipole), which is the most suitable candidate to strongly enhance the nonlinear light-matter interactions (44). Additionally, in contrast to the magnetic field, the electric field strongly enhances outside the resonator into the carved volume, as shown in the left panel of Fig. 2B, making our metasurface suitable for hybrid structure applications, where strong coupling with another material is of paramount importance. In the proposed metasurface, the contribution of other higher-order modes, such as electric quadrupole (EQ) and magnetic quadrupole (MQ), is negligible.

The scanning electron microscope (SEM) image in Fig. 2C demonstrates the high quality of the fabricated metasurfaces. The optical image in (Supplementary Materials Fig. S2), delivers comprehensive details about metasurfaces quality, as it also offers additonal insights into uniformity by examining the colour of the image and still resolves the nanostructuring. We used custom-built white light transmission spectroscopy to measure the linear transmission spectrum of metasurfaces A-C. Figure 2D displays the measured linear transmission spectrum of metasurface A. As shown in Figures 2A and 2D, a close correspondence is evident between the measured and simulated transmission spectra.

### 2.3. SHG in qBIC TMDC metasurfaces

To examine the enhancement factor (EF) of the SHG process in the qBIC TMDC metasurfaces, we employed a tunable femtosecond laser (Chameleon Ultra II and OPO, pulse width of ~200 fs) in a home-build microscopy setup. Wavelength-dependent SHG measurements have been performed on each metasurface by focusing the laser onto the metasurfaces using a 5 × microscope objective ($NA = 0.3$) to provide broader illumination area for uniform excitation. Then, the SHG signal was collected in the forward direction using a $20 \times$ microscope objective ($NA = 0.4$), enhancing SHG signal collection and spatial resolution. A short-pass filter at 800 nm has been employed to filter out the transmitted and reflected pump beams. Additionally, two waveplates have been employed to control the polarisation of the excitation beam.

The pristine WS$_2$ bulk film is 2H-type, possess inversion symmetry, leading to intrinsically weak SHG. However, the presence of interfaces can break this symmetry. This result in a non-zero, albeit weak, surface-like SHG response at the material-air and material-substrate interfaces. However, TMDCs posses high refractive index, pristine crystal quality and giant anisotropic properties (45), which are crucial for enhancing nonlinear response like SHG. To unlock this potential, we patterned pristine WS$_2$ bulk film into metasurfaces A, B, and C that support qBICs at pump wavelngths of 1220 nm, 1270 nm and 1305 nm, respectively, as depicted in Fig. 3(A). We employed above experimental setup and measured wavelength dependent SH response of our metasurfaces A, B, and C. For this type of experiment, we tune the pump wavelength to the spectral positions of the qBICs 1220 nm, 1270 nm, and 1305 nm in metasurfaces A, B, and C, respectively. We have observed significant SH enhancement compared to the unpatterned WS$_2$ film, in all three metasurfaces when pump wavelength tuned to the spectral positions of the qBICs supported by metasurfaces A, B, and C, (Supplementary Materials Sec. Nonlinear Response, Fig. S6).

Fig. 3A shows the measured transmission spectra of the three WS$_2$ metasurfaces A, B, and C, at room temperature (RT). Shaded region: Metasurface A supporting qBIC at 1220 nm (resonant at $\frac{\lambda_{qBIC}}{2}$) will interfere with the $E_0^A$ ($\lambda_{E_0^A} = 610$ nm) of WS$_2$ metaatoms, thereby achieving enormously enhanced SHG, as shown in Fig. 3B. This is the region where metasurface A is in double resonant condition, namely $\lambda_{E_0^A}$ and harmonic $\lambda_{qBIC}/2$ match or $\lambda_{qBIC} = 2 \times \lambda_{E_0^A}$, thereby significantly enhancing the SHG. Moreover, power-dependent SHG measurements confirm the nonlinear nature of the signal, exhibiting a quadratic dependence on pump power (Supplementary Materials sec. 3.2., Fig. S7), as expected for a second-order process. For metasurface B, this enhanced SHG is reduced because the doubly resonant conditon starts breaking. The SHG significantly reduced for metasurface C because $\lambda_{qBIC}/2$ falls far from the $E_0^A$ spectral position, as shown in Fig. 3A. The spectral positions of qBIC resonances can be tuned by changing the asymmetry factor in WS$_2$ arrays of metaatoms, as shown by SEM images in Fig. 3C following the color-coding schemes used in Fig. 3A.

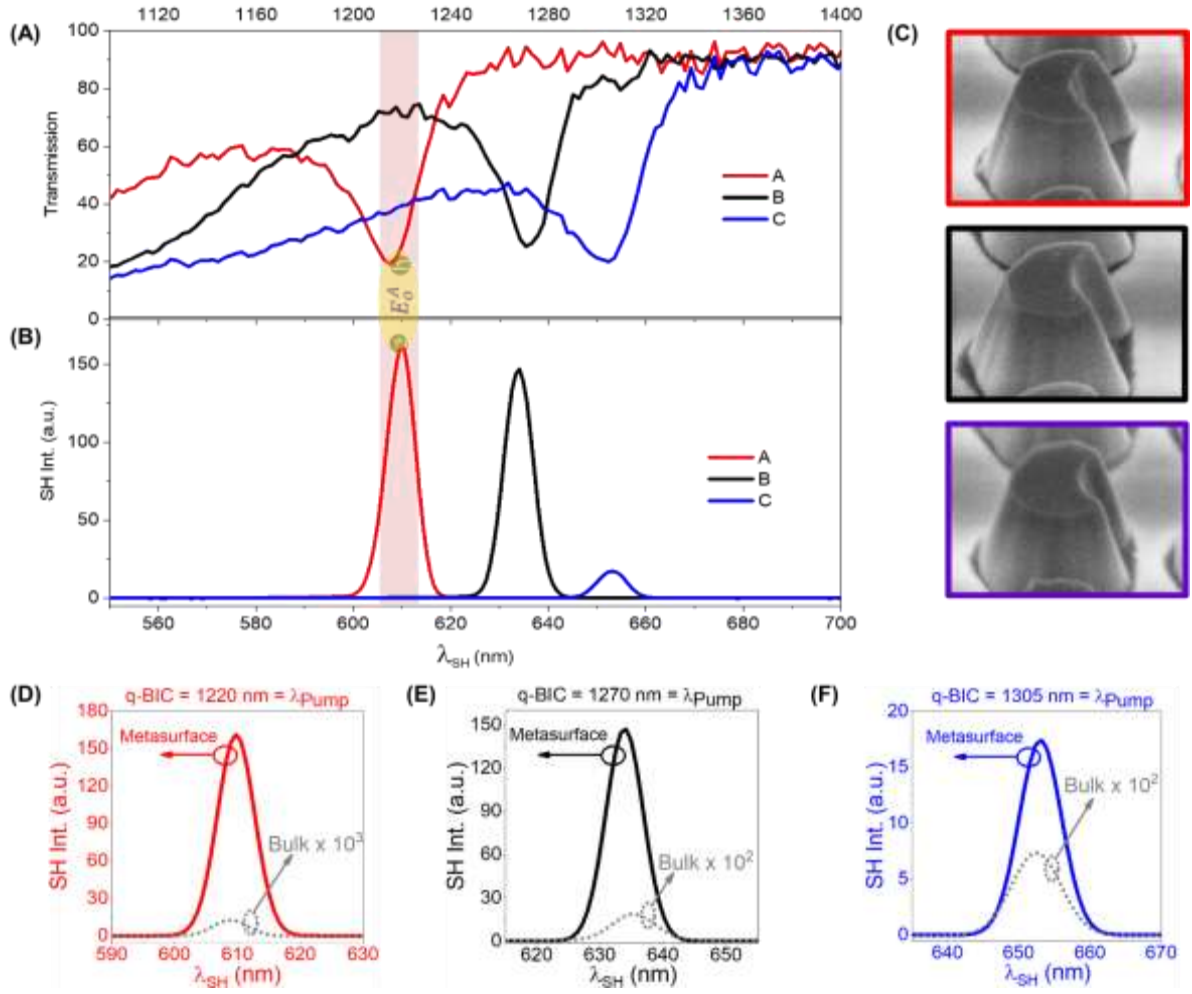

**Fig. 3. Measured SHG of qBIC WS$_2$ metasurfaces.** (A) Measured linear spectrum of metasurface A (qBIC:1220 nm), B (qBIC: 1270 nm), and C (qBIC: 1305 nm). (B) measured SHG of metasurfaces A, B, and C. Light red shaded area is the region where $\lambda_{qBIC:1220}/2$ falls at the spectral position of $E_0^A$(610 nm), and light black is the region where $\lambda_{qBIC:1270}/2$ falls at the split resonance of $E_0^A$($E_0^{*A}$: 635 nm) this split appears in the multilayer structure owing to the strong coupling between Fabry-Perot modes and $E_0^A$ (46). Moreover, light blue is the region where $\lambda_{qBIC:1305}/2$ falls at the spectral position of 652 nm, and this area is far from $E_0^A$. (C) SEM images: showing unit cells of metasurfaces A, B, and C following the color-coding schemes used in (A). (D-F) Comparison of measured SHG enhancement between metasurfaces and unpatterned WS$_2$ film: (D) Metasurface A: EF is 12,777. (E) Metasurface B: EF is 7,850. (F) Metasurface C: EF is 238.

In the shaded region, metasurface A has 9-fold stronger SHG (owing to stronger qBIC-Excitonic effects triggerd by doubly resonant condition: $\lambda_{qBIC} = 2 \times \lambda_{E_0^A}$) compared to metasurface C, as illustrated in Fig. 3B. The EF of SHG in qBIC WS$_2$ metasurfaces compared to the reminiscent bulk film (intentionally unpatterned film sitting next to metasurfaces) is presented in Fig. 3D-F. A microscope image introduced in Supplementary Materilas Fig. S2, showing the fabricated metasurfaces and unpatterned WS₂ films.

Owing to the strong excitonic-photonic interference effect in metasurface A (enabled by double resonant condition between qBIC and $E_0^A$) resulting in 4 orders of magnitude enhanced SHG compared to the unpatterned WS₂ film at the same spectral position, as shown in Fig. 3D. Notably, the enhanced SHG from metasurface A is more than 98-fold stronger than 1L-$WS_2$ (exfoliated from same bulk crystals (HQ Graphene 2H-WS₂) as we used in our metasurface) as discussed in detail in Supplementary Materials sec. 3.3. Similarly, metasurface B, for which qBIC is in the vicinity of the $E_0^A$ exciton, resulting in 3 orders of magnitude enhanced SHG compared to the unpatterned WS₂ film, as shown in Fig. 3E. For metasurface C, for which we have only qBIC resonance at pump wavelength of 1305 nm and not meeting the doubly resonant criteria ($\lambda_{qBIC} \neq 2 \times \lambda_{E_0^A}$), we observe only enhancement of 2 orders of magnitude SHG compared to the unpatterned WS₂ film, shown in Fig. 3F. The large SHG enhancement in metasurface C arises from the high-Q qBIC resonance at 1305 nm. These modes confine light efficiently, boosting the local field intensity and nonlinear response (4-5). The high-Q factor prolongs the resonance lifetime, enhancing energy buildup and SHG by 2 orders of magnitude versus unpatterned WS₂. This follows the expected quadratic scaling $I_{SH} \propto Q^2 I_o^2$ (40). In a nutshell, the qBIC alone enhances 2 orders of magnitude SHG in TMDC metasurfaces (as in metasurface C), whilst this enhancement can be boosted to 4 orders of magnitude by exploiting the excitonic-photonic interference effects in metasurface A enabled by doubly resonant criteria ($\lambda_{qBIC} = 2 \times \lambda_{E_0^A}$), as shown in Fig. 3.

### 2.4. Dynamic control on the SHG in TMDC metasurfaces

Excitonic-enhanced SHG has been previously reported in atomically thin TMDCs (28,31,32), and these excitonic effects may exist even in the bulk form (24), thereby opening opportunities to explore fundamental physics in TMDCs (47-49). Direct excitation of optical resonances at the excitonic resonance results in mode splitting (22, 48), diverting energy from the pump and introducing nonlinear loss mechanism, thereby suppressing SHG efficiency. To circumvent these limitations, we proposed a promising strategy to detune the qBIC from excitonic resonance and excite at double the excitonic wavelength ($\lambda_{qBIC} = 2 \times \lambda_{E_0^A}$). This approach leverages virtual (quantum level) exciton-qBIC interaction at RT and significantly enhance the SHG without resonant absorption losses. To dynamically control the SHG enhancement, this exciton-qBIC virtual interaction can be made tunable by three distinct approaches: by spectrally tuning the qBIC at the fundamental wavelength, by controlling/tuning the excitation of qBIC at the fundamental wavelength, or by spectrally tuning the exciton at the SH wavelength. In the first approach, we spectrally tune the qBIC at the fundamental wavelength by changing the parameter $\delta''$ of the fabricated metasurfaces (A, B, B′, and C) while keeping the other parameters constant, as illustrated in Fig. 4A. It can be observed in Fig. 4A, the metasurface A supporting qBIC at 1220 nm (meeting the doubly resonant condition: $\lambda_{qBIC} = 2 \times \lambda_{E_0^A}$) exhibits strongest SHG. As shown in Fig. 4A, this SHG enhancement starts reducing as we start breaking the doubly resonant condition by tuning the qBIC resonance towards longer pump wavelengths (1270 nm, 1290 nm, 1305 nm) via changing the parameter $\delta''$ in metasurfaces B, B′, and C, respectively. To validate the experimental approach and asses the qBIC tunability effect on exciton-qBIC double resonant condition and its related SHG enhancement, we performed full wave nonlinear simulations, as detailed in Supplementary Materials sec. 3.5.1. To varify the experimental results reported in Fig. 4(A), in our numerical simulations we spectrally tune the qBIC at pump wavelngths (via changing the asymmetry) to break and restore the double resonant condition. It can be observed in Fig. 4(A) and simulated results shown in (Supplementary Materials Fig. S10) that SHG significantly enhance when qBIC meet the double resonant condition. Both results are in good agreement.

Our second approach controls the qBIC excitation optically, simply by adjusting the incident light's polarisation angle ($\varphi$). As shown in Fig. 4B, the MD qBIC resonance in our WS$_2$ metasurface is highly polarisation dependent. For metasurface A, the qBIC at 1220 nm activates at $\varphi = 90°$ ($\lambda_{qBIC} = 2 \times \lambda_{E_0^A}$) but turns off at $\varphi = 90°$. This switching leads to 2 orders of magnitude SHG enhancement when the qBIC is on , as shown in Fig. 4B. Compared to unpatterned WS$_2$, metasurface A achieves 4 orders of magnitude SHG boost at $\varphi = 90°$, as depicted in Fig. 3D, making polarisation a powerful tuning knob for SHG at RT. Polarisation-resolved SHG measurements ($\varphi = 0°-360°$) reveal a dipole-like pattern (Supplementary Fig. S8), contrasting with bulk WS$_2$'s six-fold symmetry. This shows our qBIC metasurface can optically reshape SHG emission-a unique feature controllable purely by $\varphi$.

The final approach employs spectral tuning the exciton $E_0^A$ at SH wavelengths by leveraging some external influences like temperature, to align with the desired state. It is noteworthy that this tunability can also be achieved via other external stimuli, such as strain (26) or electric field (28), but temperature was the accessible method in our current experimental setup. In our work, to clearly indicate the SHG enhancement factor at RT in comparison to the other temperatures, we have considered small temperature change of $-100°$ to $100°$ around RT and such temperature range can spectrally tune $E_0^A$ at SH wavelengths (from 600 nm to 622 nm) to break and/or restore the doubly in resonant condition ($\lambda_{qBIC} = 2 \times \lambda_{E_0^A}$), as shown in Fig. 4C. To experimentally examine the temperature-dependent SHG behaviour of our targeted metasurface A, we placed our sample in Linkam chamber (attached to the temperature controller and liquid nitrogen), varied the temperature of our fabricated metasurface A and then measured its wavelength-dependent SHG in the temperature range $-100°$ to $100°$ with the following steps ($-100°, -50°,$ (RT), $100°$), as shown in Fig. 4C. It can be observed in Fig. 4(C) that SHG at RT is almost 4.3x stronger than the SHG at other temperatures i.e $-100°$. This is because at RT the $E_0^A$ qBIC are meeting the doubly resonance condition: $\lambda_{qBIC} = 2 \times \lambda_{E_0^A}$. However, at other temperatures the $E_0^A$ starts spectrally detuning from its RT spectral position (~610 $nm$) and exhibits a condition where $\lambda_{E_0^A} \neq \frac{\lambda_{qBIC}}{2}$. Owing to the wider nature of $E_0^A$ resonance in bulk WS$_2$, this shift is not sufficient to completely break the doubly resonant condition but is enough to affect the SHG enhancement and observe its effect.

To verify our experimental results shown in Fig. 4(C), we further performed full-wave numerical simulations to assess the temperature tunability of excitonic-photonic enhanced SHG from metasurface A, and results are illustrated in Fig. 4(D). The details of numerical simulations related to the temperature tunability can be found in (Supplementary Materials under section 3.5.2 and Fig. S13). In Fig. 4(D), the blue dashed line is the eye guide for the spectral position of $E_0^A$, whilst the grey dashed dot line represents the spectral position of qBIC in proposed WS$_2$ metasurface A. It can be observed in Fig. 4(D) that variations in temperature relative to the RT spectrally detune the $E_0^A$ from its spectral position at RT (~610 $nm$). This spectral tuning of $E_0^A$ as a function of temperature can break and subsequently restore the doubly resonant condition ($\lambda_{qBIC} = 2 \times \lambda_{E_0^A}$) established at RT in metasurface A, thereby enabling dynamic control on the SHG enhancement as a function of an external stimulus, as illustrated in Fig. 4(D). The measured and simulated results illustrated in Fig. 4(C) and (D), respectively, are in good agreement and show that SHG enhances at RT when $E_0^A$ and qBIC satisfy the doubly resonant condition ($\lambda_{qBIC} = 2 \times \lambda_{E_0^A}$). The underlying physics of the doubly resonant condition enhanced SHG along with its dependence on qBIC spectral position as function of $\delta''$ and external stimuli such as temperature, can be comprehensively explained via the application of Fermi's Golden rule for SHG conversion efficiency (for details see Supplementary Materials section 3.5.3).

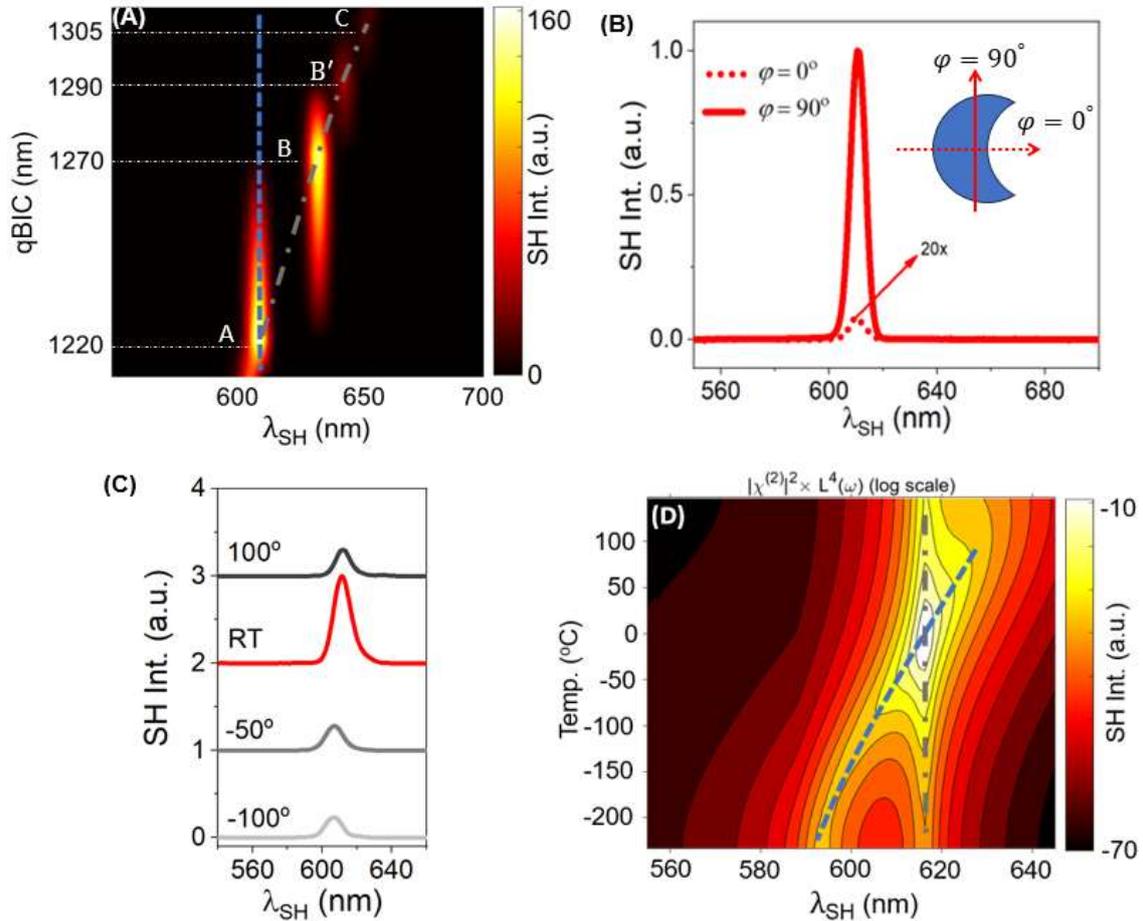

**Fig. 4. Measured and simulated tunable SHG of qBIC WS$_2$ metasurfaces.** (**A**) Measured SHG mapping as a function of qBIC and excitation wavelength. The spectral position position of qBIC has been tuned by changing the $\boldsymbol{\delta''}$. The vertical dashed blue line shows the position of $\mathbf{E_0^A}$ in bulk WS$_2$, while the grey dash-dot line shows the spectral tuning of qBIC as a function of the parameter $\boldsymbol{\delta''}$. The horizontal dashed lines are the eye guide for the qBIC spectral positions of the metasurfaces A, B, **B′**, and C. (**B**) Measured SHG enhancement of metasurface A when the pump polarisation angle $\boldsymbol{\varphi} = \mathbf{0}°$ and $\boldsymbol{\varphi} = \mathbf{90}°$, the inset showing the schematic of $\boldsymbol{\varphi}$ of the incident light with respect to the crescent shaped metaatoms geometry. (**C**) Measured wavelength-dependent SHG of metasurface A as a function of temperature with temperature change of $\mathbf{100}°$ **to** $-\mathbf{100}°$ around RT. (**D**) Simulated wavelength-dependent SHG of metasurface A as a function of temperature ranging from $\mathbf{100}°$ to $-\mathbf{190}°$. Results presented in (**A-B**) are measured at ambient conditions (RT).

## 3. Discussion

In contrast to artificially stacked hybrid systems (monolayer coupled to other dielectric metasurfaces (38,50,51), bulk TMDCs naturally provide a cleaner system and an exciting opportunity to explore strong interactions between inelastic (excitonic properties) and elastic scattering properties (such as high-Q qBICs) (46,48,52). Owing to many exciting properties (24,48,53), bulk TMDCs offer the most promising avenue to date to subwavelength solid-state devices for tunable nonlinear optics (39). In this work, we have reported all optically and thermally tunable giant SHG enhancement in single-crystalline qBIC $WS_2$ metasurfaces. The giant SHG intensity enhancement in $WS_2$ metasurfaces is the result of the simultaneous resonant conditions or doubly resonant criteria ($\lambda_{qBIC} = 2 \times \lambda_{E_0^A}$), between $E_0^A$ and qBIC at SH and fundamental wavelengths.

To enhance the SH intensity, three approaches have been explored: spectral tuning of qBIC at pump wavelength as a function of the asymmetry parameter $\delta''$, tuning excitation of induced qBIC at particular pump wavelength as function of $\varphi$ of the incident light, and spectral tuning of $E_0^A$ at SH wavelength as a function of temperature. In comparison to the change in geometrical parameters of metasurfaces, the last two approaches provide us easy platforms to manipulate the doubly resonant condition of $E_0^A$ and qBIC as a function of $\varphi$ of the incident light and temperature to control the SHG in TMDC metasurfaces.

To engineer the double resonant condition in metasurface A, we controlled the excitation of qBIC at pump wavelength of 1220 nm as a function of $\varphi$ of the incident light. By just changing the $\varphi$ of the incident light from $90°$ to $0°$, we can activate or deactivate qBIC resonance at pump wavelength of 1220 nm to restore and/or break the doubly resonant condition in metasurface A. When the qBIC is activated ($\varphi = 90°$), the metasurface A exhibits 2 orders of magnitude SHG enhancement in comparison to the qBIC deactivation state ($\varphi = 0°$). This all optically controlled method also helped to modulate the intrinsic six-fold SHG radiation pattern of bulk $WS_2$ metasurface into dipole pattern. Moreover, we performed wavelength-dependent SHG at temperature range of $-100°$ to $100°$ around RT, to further engineer the doubly resonant condition in metasurface A. By increasing and decreasing the temperature, the spectral position of $E_0^A$ can be tuned towards longer and shorter wavelengths, respectively. Therefore, by changing the temperature, it can be observed that when the harmonic energy of qBIC of metasurface A overlaps with the energy of $E_0^A$ (satisfying the double resonant condition), then, a giant enhancement of 4 orders of magnitude in SHG can be observed compared to unpatterned

WS$_2$ film. Interestingly, in doubly resonant condition at RT, the proposed WS$_2$ crescent metasurface A achieves SHG efficiency 98-fold (experimentaly) and 3 orders of magnitude (numerically) stronger than that of 1L-WS$_2$. The difference in experimental and simulated results is due to the difference in the measured and simulated Q factor of qBIC and local pump intensity. This substantial enhancement underscores the unique advantage of exploiting qBIC resonances alongwith the intrinsic excitonic resonances in bulk TMDC metasurfaces for nonlinear light-matter interactions. We also show that 9-fold enhancement of metasurface A compared to the metasurface C, for which $E_0^A$ and harmonic of qBIC are not in double resonant condition. In metasurface A, the highest measured SHG efficiency is estimated about $5.8 \times 10^{-9}$ at incident peak power of $3.56\, kW$ (Supplementary Materials Sec. 3.6). The simulated SHG efficiency of metasurface A is 2 orders of magnitude stronger than the 1L-WS$_2$ and 5 orders of magnitude stronger than the unpatterned WS$_2$ (sharing the parent flake as of metasurfaces).

The enhanced SHG from our qBIC metasurfaces can be applied to a range of practical applications, including frequency conversion, sensing, and integrated photonic circuits (21, 54-55). For frequency conversion, the significant SHG enhancement could enable more efficient generation of tunable wavelengths for telecom, spectroscopy, and biomedical imaging. In sensing, this enhanced SHG could improve optical sensor sensitivity and detection limits for nonlinear optical sensors, allowing for precise distance measurements with nanometer resolution, aiding in chemical, biological, and environmental sensing. Additionally, our metasurfaces could be integrated into photonic circuits for on-chip frequency conversion, enabling more compact and efficient devices for quantum information processing. By detecting changes in the SHG signal, our metasurfaces can identify the presence of specific biomolecules, making them valuable for biomedical diagnostics. Localization of light via magnetic type resonances in dielectric metasurfaces can suppress undesirable free carrier contribution, leading to ultrafast all optical switching (21). As our metasurfaces support pure magnetic type qBICs, thereby making them valuable candidate for ultrafast all-optical switching applications. The proposed method is not only the potential for tunable SHG enhancement but also lay foundation for extending this approach to advanced nonlinear optical processes, such as spontaneous parametric down-conversion (56) for next-gen quantum applications. Last but not least, the proposed metasurface A achieves 4 orders of magnitude SHG enhancement over unpatterned bulk WS$_2$ and enables $10^2 \times$ all-optical modulation-advances critical for on-chip nonlinear optics.

We note that electrical tuning of the excitonic resonances can also be employed to realize compact tunable devices. However, the mixing between exciton and trion resonances must be taken into account.

**Acknowledgements**:

This work was supported by Muhammad Iqbal and The Punjab Educational Endowment Fund (PEEF) Pakistan, by the Australian Research Council through Centres of Excellence (CE200100010, CE170100012). We acknowledge the use of the ACT node of the Australian National Fabrication Facility (ANFF). MR acknowledges support from the UK Research and Innovation Future Leaders Fellowship (MR/T040513/1).

**Methods**

**Device Fabrication.** We fabricated multiple arrays of $WS_2$ crescent-shaped metaatoms ($WS_2$ metasurfaces) on a sapphire ($Al_2O_3$) substrate by employing electron-beam lithography (EBL) method (57,58). To ensure pristine sample quality, $WS_2$ van der Waals (vdW) material was mechanically exfoliated onto the substrate. The thickness of the transferred flakes was measured using a surface profilometer. Following exfoliation, the flakes were cleaned and coated with a single layer of ZEP positive resist, followed by a thin layer of conductive polymer to mitigate charging effects during EBL. The selected flake was patterned into crescent-shaped metaatoms using EBL. Post-development, a 60 nm thick aluminum layer was deposited via electron-beam evaporation to serve as a dry etching mask. Crescent-shaped metaatoms were then defined by dry etching using an inductively coupled plasma process with $SF_6$, $CHF_3$, and Ar gases. The aluminum caps atop the truncated-cone pillars were subsequently removed using a lift-off process. Finally, the quality and dimensions of the $WS_2$ metasurfaces were assessed through scanning electron microscopy (SEM).